    \documentclass[a4paper,9pt,twocolumn]{extarticle}
    \usepackage[utf8]{inputenc}
    \usepackage[T1]{fontenc}
    \usepackage[margin=2cm,columnsep=0.6cm]{geometry}
    \usepackage{amsmath,amssymb}
    
    \usepackage{graphicx}

    \usepackage[font=small,labelfont=bf]{caption}
    \usepackage{xcolor}
    \usepackage{authblk}
    \usepackage[colorlinks=true, citecolor=black, linkcolor=black, urlcolor=blue]{hyperref}

    \newcommand{\keywordsline}[1]{\vspace{0.5em}\noindent\textbf{Keywords:} #1}
    \usepackage{environ}
    \RenewEnviron{abstract}{\global\let\theabstract\BODY}

\newcommand{\papertitle}{Modeling elasto-viscoplastic free-surface flows with different yield surfaces}

\newcommand{\keywordone}{Elasto-viscoplastic flows}
\newcommand{\keywordtwo}{Free surface flows}
\newcommand{\keywordthree}{Cohesive flows}
\newcommand{\keywordfour}{Overstress model}
\newcommand{\keywordfive}{Yield-stress fluid}
\newcommand{\keywordsix}{Yield surfaces}
\newcommand{\keywordseven}{Finite strain}
\newcommand{\keywordeight}{Material point method}

\usepackage[authoryear]{natbib}
\usepackage{orcidlink}
\usepackage{bm} 
\DeclareMathOperator*{\dev}{dev}
\DeclareMathOperator*{\tr}{tr}

\begin{document}

    \title{\huge{\papertitle}}
    \author[1,2]{\orcidlink{0000-0002-2016-7233} Lars Blatny\textsuperscript{*}}
    \author[3]{\orcidlink{0009-0006-6554-8087} Alexandre Pellet}
    \affil[1]{Institut Jean Le Rond d'Alembert, Sorbonne Universit\'e, CNRS - UMR 7190, F-75005 Paris, France}
    \affil[2]{Institute for Geotechnical Engineering, ETH Zürich, CH-8093 Zürich, Switzerland}
    \affil[3]{Universit\'e Grenoble Alpes, INRAE, CNRS, IRD, Grenoble INP, IGE, F-38000 Grenoble, France}
    \date{\small{\today}}

\begin{abstract}
Elasto-viscoplasticity provides a unified way of describing yield-stress fluids which may exhibit both solid-like and fluid-like behavior. 
In this work, we present a finite strain overstress-type elasto-viscoplastic framework designed to facilitate the incorporation of different yield surfaces. 
Within this framework, we compare several yield-surface choices and assess the associated challenges. 
We consider three representative yield surfaces: (i)~pressure-independent, (ii)~pressure-sensitive frictional and (iii)~capped surfaces, corresponding to von Mises, Drucker--Prager, and modified Cam--clay models, respectively. 
In the case of von Mises, the proposed formulation naturally recovers the well-known Bingham and Herschel--Bulkley rheologies which are characterized by a single critical yield stress. 
We discuss in detail the singularity of the Drucker--Prager yield surface which requires a special treatment.
In particular, we show that the modified Cam--clay model can be used to conveniently circumvent this singularity under the right conditions, retrieving the expected solution of Drucker--Prager.
Implemented within a hybrid Eulerian--Lagrangian scheme, the general framework presented here enables efficient simulations of elasto-viscoplastic flows in two or three spatial dimensions, not requiring regularizing the solid-fluid transition nor a separate free-surface treatment. 
Numerical benchmark simulations illustrate how yield surface geometry affects velocity profiles, plug formation and compressibility. 
\end{abstract}

    \makeatletter
    \twocolumn[
      \begin{@twocolumnfalse}
        \maketitle
        \vspace{-2em}
        \begin{center}\bfseries Abstract\end{center}
        \begin{quote}\normalsize
            \theabstract
        \end{quote}
        \begin{quote}\small
            \keywordsline{\keywordone; \keywordtwo; \keywordthree; \keywordfour; \keywordfive; \keywordsix; \keywordseven; \keywordeight }
        \end{quote}
        \vspace{1.5em}
      \end{@twocolumnfalse}
      ]
      \makeatletter
        \renewcommand\thefootnote{\fnsymbol{footnote}}
        \footnotetext[1]{Corresponding author: \texttt{larsblatny@gmail.com}}
        \renewcommand\thefootnote{\arabic{footnote}}
        \setcounter{footnote}{0}
        \makeatother
    \makeatother

\section{Introduction}
Yield‑stress fluids are ubiquitous in natural and industrial contexts. 
Examples include food products such as ketchup, personal care products like toothpaste and gels, as well as natural materials like mud and clay.
Such materials resist deformation up to a critical stress threshold, below which they behave like solids and above which they flow like viscous fluids. 
This behavior is generally attributed to microstructural rearrangements, with transitions between jammed and unjammed states~\citep{COUSSOTavalanche2002, NGOUAMBAelastoplastic2019}. 
Recent reviews by \cite{BALMFORTHimplications2025, COUSSOTfifty2025, SARAMITOprogress2017, BONNyield2017, BALMFORTHyielding2014, COUSSOTyield2014} describe and discuss yield-stress fluids, and readers are referred to these excellent texts for further background on the topic. 

The classical viscoplastic model attributed to Bingham~\citep{BINGHAMfluidity1922}  introduces a yield stress into a Newtonian framework. 
This is further extended by the Herschel–Bulkley model~\citep{HERSCHELkonsistenzmessungen1926} which introduces an extra parameter controlling shear-thinning or shear-thickening. 
Another example of a yield-stress viscoplastic model is the~$\mu(I)$-rheology, a now widely popular description of dense granular flows~\citep{GDRMIDIdense2004, JOPconstitutive2006}, which has been implemented in various Navier-Stokes solvers (e.g., \citet{LAGREEgranular2011, IONESCUviscoplastic2015, RAUTERgranular2020}). 
However, attributed in part to the non-smooth transition between flow and no flow by a sharp threshold, these well-known viscoplastic models should be regularized or augmented in numerical solvers \citep{PAPANASTASIOUflows1987, BLACKERYcreeping1997, FRIGAARDusage2005, SARAMITOprogress2017, BARKERpartial2017}. 
Moreover, such purely viscoplastic descriptions completely ignore elasticity, which not only governs the material response below yielding, but can also influence the flow dynamics (see, e.g., \cite{JOSSICflow2013, FRAGGEDAKISyielding2016a, LOPEZrising2018}). 
This recognition has motivated the development of elasto‑viscoplastic~(EVP) constitutive models, which combine elastic deformation and rate‑dependent (viscous) plastic flow within a unified model.

In the non-Newtonian fluid mechanics community, Pierre Saramito~\citep{SARAMITOnew2007, SARAMITOnew2009, SARAMITOcomplex2016, SARAMITOnew2021, SARAMITOanother2026} has made significant contributions to the development of modern~EVP models. 
In his 2007 work, \cite{SARAMITOnew2007} introduced an~EVP constitutive model that integrates elasticity with viscoplastic flow, providing a thermodynamically consistent description of yield‑stress fluids that behave as solids below yield and that flows above yield. 
Later, this was extended to incorporate Herschel–Bulkley viscoplasticity~\citep{SARAMITOnew2009}. 
This model has been implemented in the finite element method~\citep{CHEDDADInew2013} and more recently in the finite volume method~\citep{SYRAKOSfinite2020}.
It has been thoroughly benchmarked against alternative viscosity regularization methods and experimental data by~\citet{FRAGGEDAKISyielding2016}, demonstrating great promise of this constitutive framework, although certain limitations have also been very recently identified~\citep{SHEMILTstartup2026, SHEMILTtransition2026}. 
Subsequent developments explored brittle~EVP behavior with a pressure‑sensitive yield surface~\citep{SARAMITOnew2021}, and very recent efforts, concurrent with the work presented here, have refined the thermodynamic environment of~EVP models~\citep{SARAMITOanother2026}. 

As strongly indicated by~\cite{COUSSOTslow2018}, yield-stress fluids are, at least in the slower limit, essentially plastic flow under large deformation, and thus the field of plasticity can benefit the constitutive and computational development of yield-stress fluids. 
This perspective suggests that concepts developed in finite strain plasticity can provide a natural framework for the modeling of such fluids.
Although numerical modeling of materials with complicated yield criteria is well understood in solid mechanics, these established methods have not always been directly transferrable to flow problems.
In granular flow modeling, some studies have introduced different yield surfaces into the $\mu(I)$-rheology in both a viscoplastic~\citep{BARKERwellposed2017, SCHAEFFERconstitutive2019, RAUTERgranular2020} and elasto-viscoplastic~\citep{DUNATUNGAcontinuum2015, BLATNYcritical2024} setting. 
Recognizing that the $\mu(I)$-rheology can be interpreted as a purely viscoplastic Drucker–-Prager model with variable viscosity,~\cite{IONESCUviscoplastic2015} proposed a simplified viscoplastic Drucker–-Prager formulation with constant viscosity, which also showed agreement with experimental observations of granular flows. 
In the context of landslide flow modeling, \cite{PASTORdepth2015, PASTORviscoplastic2015} considered the combination of Perzyna viscoplasticity with different yield surfaces in an infinite EVP landslide model. 
This concept has been a source of motivation for the present work, in which we propose a general three-dimensional formulation and provide a verification and discussion of the resulting solutions, which has been outside the scope of the previous geophysical flows-focused studies. 

Rather than focusing on a particular material rheology, the purpose of the present work is to compare and assess the challenges associated with three different yield surfaces, all well-known from solid mechanics, for the modeling of yield-stress flow problems.
We address this by formulating a finite strain overstress‑type~EVP model within a hybrid Eulerian–Lagrangian numerical model that can capture local transitions between solid- and fluid-like behavior, allowing two- and three‑dimensional simulations of free‑surface flows. 
In particular, the material point method~(MPM) is used as the numerical scheme capable of handling large deformation and flow of~EVP materials.

\section{Theory}

\subsection{Finite strain elastoplasticity}
\label{sec:yield}
We rely on a finite strain framework with a multiplicative decomposition of the deformation gradient $\bm{F} = \bm{F}^E \bm{F}^P$ following \citet{LEEelasticplastic1969}. 
From the deformation gradient, different strain measures can be defined, e.g., the Hencky strain tensor~$\bm{\varepsilon}$ defined as
\begin{equation}
\bm{\varepsilon} 
= \frac{1}{2} \ln \bm{F} \bm{F}^T
=  \sum\limits_{\alpha=1}^{d} \ln \lambda_\alpha \text{ } \bm{n}_\alpha \otimes \bm{n}_\alpha 
\label{hencky}
\end{equation}
where $d$ denotes the number of spatial dimensions, the unit vectors $\bm{n}_\alpha$, $\alpha = 1,...,d$, define the spatial principal directions and $\lambda_\alpha$, $\alpha = 1,...,d$, are the principal streches, i.e., the  singular values of $\bm{F}$.
For simplicity, all models presented here will be based on Hencky's elasticity model which relates the Kirchhoff stress~$\bm{\sigma} = \det(\bm{F}) \bm{\sigma}_*$, where $\bm{\sigma}_*$ is the Cauchy stress, to the elastic Hencky strain~$\bm{\varepsilon}^E$ with a form familiar from isotropic linear elasticity, in particular,
\begin{equation}
	\bm{\sigma} = \mathcal{C} : \bm{\varepsilon}^E =  \Lambda \tr\big(\bm{\varepsilon}^E\big) \bm{I} + 2 G \bm{\varepsilon}^E \text{,}
	\label{henckyelasticity}
\end{equation}
where $\Lambda$ and $G$ are the two Lam\'e parameters. 
These two parameters may be related to the perhaps more familiar Young's modulus~$E = \frac{G(3\Lambda+2G)}{\Lambda+G}$ and Poisson's ratio~$\nu = \frac{\Lambda}{2(\Lambda + G)}$, here assuming three dimensions.
In addition, the bulk modulus can be defined as $K=\Lambda + \frac{2}{d}G$. 
The properties of the finite strain hyperelastic model given by Eq.~\eqref{henckyelasticity} and its relation to other models are given in, e.g., \cite{BRUHNSselfconsistent1999, XIAOhenckys2002}.

A yield function $y=y(\bm{\sigma})$ determines the onset of plastic deformations at $y=0$, with elastic stress states satisfying $y<0$.  
The velocity gradient can be additively decomposed as
\begin{equation}
\nabla \bm{v} = \bm{l}^E + \bm{l}^P
\end{equation}
and a plastic flow rule imposes 
\begin{equation}
    \bm{l^P} = \dot{\bar{\gamma}} \frac{\partial g}{\partial \bm{\sigma}}
    \label{plasticvelgrad}
\end{equation}
where $g=g(\bm{\sigma})$ is known as a plastic potential, defining the direction of plastic flow, and the scalar $\dot{\bar{\gamma}}$ is known as the plastic multiplier. 
The flow rule must induce a non-negative plastic rate of dissipation, i.e., $\bm{\sigma} : \bm{l}^P \geq  0$. 
The choice $g = y$ is called an associative flow rule and is a consequence of the principle of maximum plastic dissipation. 
In particular, it can be shown that maximizing $\bm{\sigma} : \bm{l}^P$ subject to $y \leq 0$ leads to $\bm{l^P} = \dot{\bar{\gamma}} \frac{\partial y}{\partial \bm{\sigma}}$
where $\dot{\bar{\gamma}}$ can be interpreted as a Lagrange multiplier \citep{SIMOframework1988, SIMOalgorithms1992}.

In this work, $y$ and $g$ will be chosen as functions of the principal stresses
\begin{equation}
p = -\frac{1}{d} \tr(\bm{\sigma)} 
\text{ \ and \ }
\tau = \frac{1}{\sqrt{2}} || \dev(\bm{\sigma}) ||
\label{p_and_q}
\end{equation}
where $p$ is termed the isotropic pressure and~$\tau$ the equivalent shear stress.
Here, we used the notation~$|| \bm{A} || = \sqrt{\bm{A} : \bm{A}} = \sqrt{\tr(\bm{AA})}$ and~$\dev(\bm{A}) = \bm{A} - \frac{1}{d}\tr(\bm{A})\bm{I}$ for a symmetric second-order tensor $\bm{A}$. 
The particular form of Eq.~\eqref{p_and_q} is chosen such that the two-dimensional case $d=2$ provides plane strain conditions with the assumption $\nu=0.5$ for yield and plastic flow, see, e.g., \citet{TIANratio2009, WOJCIECHOWSKInote2018} and references therein. 

\begin{figure}
    \centering
    \includegraphics[width=0.7\linewidth]{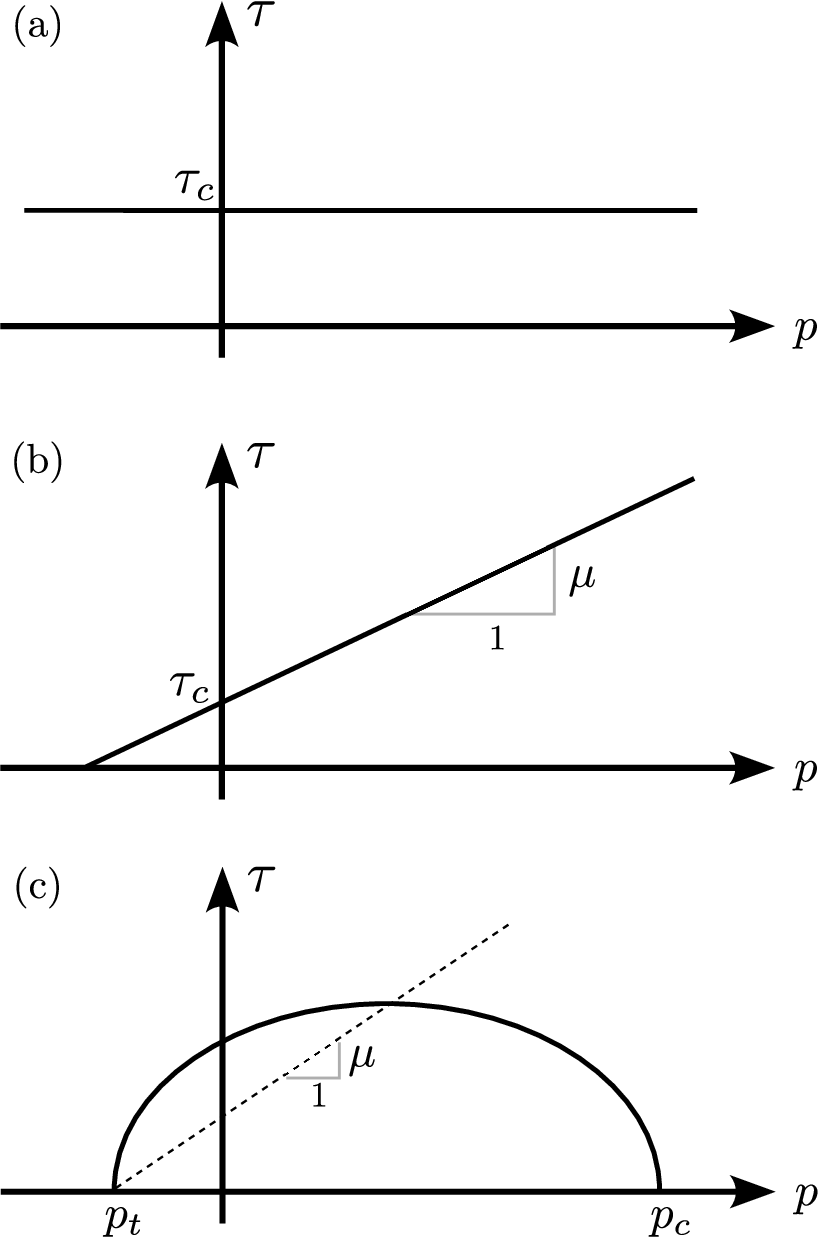}
    \caption{
      Yield surfaces: (a) von Mises, (b) Drucker--Prager and (c) modified Cam--clay.
    }
    \label{fig:yields}
\end{figure}

In the remainder of this section, three plastic models are outlined; von Mises, Drucker--Prager and modified Cam--clay. 
Their yield function can all be written on the general form
\begin{equation}
y(p,\tau) = \tau - \tau_y(p)
\label{yield}
\end{equation}
where $\tau_y(p)$ is a yield stress that may depend on $p$.
The different yield surfaces $y=0$ are sketched in Fig.~\ref{fig:yields}.

In the von Mises model~\citep{VONMISESmechanik1913} the yield strength does not depend on pressure, i.e.,
\begin{equation}
\tau_y(p) = \tau_c
\label{VM}
\end{equation}
where $\tau_c>0$ is a constant (assuming no hardening).

The Drucker--Prager model \citep{DRUCKERSoil1952} yield stress given by
\begin{equation}
\tau_y(p) = \mu p + \tau_c 
\label{DP}
\end{equation}
where $\mu>0$ is the internal friction and $\tau_c \geq 0$ is a cohesive stress. 
With an associative flow rule, $g=y$, this would, for any plastic deformation, imply a volume increase, i.e., $\tr(\bm{l}^P) > 0$. 
Consequently, this model is often used with a von Mises plastic potential, i.e., $g=\tau$. 

The modified Cam--clay model~\citep{ROSCOEgeneralized1968} is characterized by an evolving yield surface of ellipsoidal shape in principal stress space, in particular,
\begin{equation}
\tau_y(p) =\mu \sqrt{ (p-p_t) (p_c-p) } \text{,}
\label{MCC}
\end{equation}
where $\mu>0$ defines the so-called critical state line, $p_c \geq 0$ is an isotropic compressive strength and $p_t \leq 0$ is an isotropic tensile strength. 
Following \cite{GAUMEdynamic2018}, we define $\beta \geq 0$ as a dimensionless measure of cohesion such that $p_t = -\beta p_c$.
Based on critical state soil mechanics \citep{SCHOFIELDcritical1968}, this model is conceived from the idea that there exists a critical state where the material can continue to shear without further changes to its volume or stress. 
This model is usually used with an associative flow rule, thus promoting the possibility of both dilation and compaction (until critical state is reached) and a hardening law based on the accumulated plastic volumetric deformation $\varepsilon_V^P = \int \tr(\bm{l}^P) \ dt$. 
As such, the critical state must be the stress state at the apex of the ellipsoidal yield surface, which will always be along the critical state line defined by $\mu$ (see Fig.~\ref{fig:yields}c).
The hardening law considered here is given by
\begin{equation}
    p_c(\varepsilon_V^P) = p_c^0 e^{-\xi \varepsilon_V^P}
    \label{hardening_exp}
\end{equation}
where $\xi>0$ is a hardening parameter and $p_c^0$ is an initial compressive strength~\citep{BLATNYcritical2024}.

\subsection{The overstress approach}
\label{sec:overstress}
In rate-independent elastoplasticity, the Kuhn--Tucker conditions must be satisfied, i.e.,
$\dot{\bar{\gamma}} \geq 0$, $y \leq 0$ and $\dot{\bar{\gamma}} y = 0$. 
This ensures that plastic states can only exist on the yield surface $y=0$.
In the overstress approach to viscoplasticity, plastic states may satisfy $y > 0$ and these conditions no longer hold. 
Prominently, the overstress model of \cite{PERZYNAconstitutive1963, PERZYNAfundamental1966} can be written in a general way as 
\begin{equation}
    \bm{l}^P =
    \begin{cases}
        \frac{\Psi(y)}{t_v} \bm{n} , & \text{if } y > 0 \\
        \bm{0},               & \text{otherwise}
    \end{cases}
    \label{Perzyna_general}
\end{equation}
where $\Psi(y)$ is a scalar overstress function and $\bm{n} = \frac{\partial g / \partial \bm{\sigma}}{|| \partial g / \partial \bm{\sigma} ||}$ determines the flow rule. 
The overstress approach stands in contrast to, e.g., the consistency approach~\citep{WANGviscoplasticity1997}, where, instead of relying on the rate-independent yield function $y(\bm{\sigma})$, one formulates a rate-dependent yield function such that the Kuhn--Tucker conditions can still be enforced. 
Interestingly, this can be shown to be essentially equivalent to the Perzyna approach~\citep{HEEREScomparison2002}.

A popular special case of Eq.~\eqref{Perzyna_general} is found by considering a von Mises plastic potential, i.e., $g=\tau$ and yield surfaces described by a (potentially pressure-dependent) shear stress $\tau_y(p)$ as outlined in Section~\ref{sec:yield},
\begin{equation}
    \dot{\gamma} =
    \begin{cases}
        \frac{1}{t_v} \Big( \frac{\tau-\tau_y}{\tau_y} \Big)^{1/s}, & \text{if } \tau > \tau_y \\
        0,               & \text{otherwise}
    \end{cases}
    \label{Perzyna}
\end{equation}
where $\dot{\gamma} = \sqrt{2} || \dev(\bm{l}^P)||$ is denoted the equivalent plastic shear strain rate, following the decomposition $\bm{l}^P = \frac{1}{d} \tr(\bm{l}^P) \bm{I} + || \dev(\bm{l}^P)|| \frac {\dev(\bm{\sigma}) }{|| \dev(\bm{\sigma}) ||}$.
In Eq.~\eqref{Perzyna}, the two viscous parameters are the viscous time~$t_v$ and the viscous exponent~$s$. 

\cite{PERICclass1993} proposed the following slight modification, 
\begin{equation}
    \dot{\gamma} =
    \begin{cases}
        \frac{1}{t_v} \Big( \Big(\frac{\tau}{\tau_y} \Big)^{1/s} -1 \Big), & \text{if } \tau > \tau_y \\
        0,               & \text{otherwise}
    \end{cases}
    \label{Peric}
\end{equation}
which, unlike the previous formulation, recovers the rate-independent solution as $s \rightarrow 0$ in addition to $t_v \rightarrow 0$. This will be shown in Section~\ref{sec:numerics}

The overstress model typically attributed to \citet{DUVAUTinequations1972} is often given by the general form
$\bm{\sigma} - \mathcal{P}\bm{\sigma} = t_v \ \mathcal{C} : \bm{l}^P$ where $\mathcal{P}\bm{\sigma}$ is a projection of the stress onto the yield surface~\citep{SIMOcomputational1998}. 
Typically, the latter is a closest point projection in stress space in the sense of an associative plastic flow rule.
This Duvaut--Lions model can be rewritten as
\begin{equation}
    \bm{l}^P = 
    \begin{cases}
        \frac{1}{t_v} \ \mathcal{C}^{-1} : (\bm{\sigma}-\mathcal{P}\bm{\sigma}), & \text{if } y > 0 \\
        \bm{0},               & \text{otherwise}
    \end{cases}
    \label{DL}
\end{equation}
In the special case of a von Mises plastic potential and yield surfaces which can be described by a (pressure-dependent) shear stress $\tau_y(p)$ as outlined in Section~\ref{sec:yield}, this can be reduced to 
\begin{equation}
    \dot{\gamma} =
    \begin{cases}
        \frac{1}{t_v} \frac{\tau-\tau_y}{G}, & \text{if } \tau > \tau_y \\
        0,               & \text{otherwise}
    \end{cases}
    \label{DL_G}
\end{equation}
which, when compared to Eq.~\eqref{Perzyna}, motivates a generalized expression 
\begin{equation}
    \dot{\gamma} =
    \begin{cases}
        \frac{1}{t_v} \Big( \frac{\tau-\tau_y}{G} \Big)^{1/s}, & \text{if } \tau > \tau_y \\
        0,               & \text{otherwise}
    \end{cases}
    \label{DL_G_s}
\end{equation}
or, even more generally, a generalized Duvaut--Lions formulation can be proposed as
\begin{equation}
    \bm{l}^P = 
    \begin{cases}
        \frac{1}{t_v} \ \big( \mathcal{C}^{-1} : (\bm{\sigma}-\mathcal{P}\bm{\sigma}) \big)^{1/s}, & \text{if } y > 0 \\
        \bm{0},               & \text{otherwise}
    \end{cases}
    \label{DL_general}
\end{equation}
This demonstrates the equivalence between the Perzyna and Duvaut--Lions formulations. 
From the point of view of Perzyna, if 1) the normalization constant (i.e., the denominator) in the overstress function $\Psi$ is chosen appropriately (e.g., in the example above as $G$ instead of $\tau_y$) and 2) the plastic potential $g$ is chosen in line with the projection operator $\mathcal{P}$, they are equivalent.

\subsection{Steady-state velocity profiles}
\label{sec:profiles}

\begin{figure}
    \centering
    \includegraphics[width=0.5\linewidth]{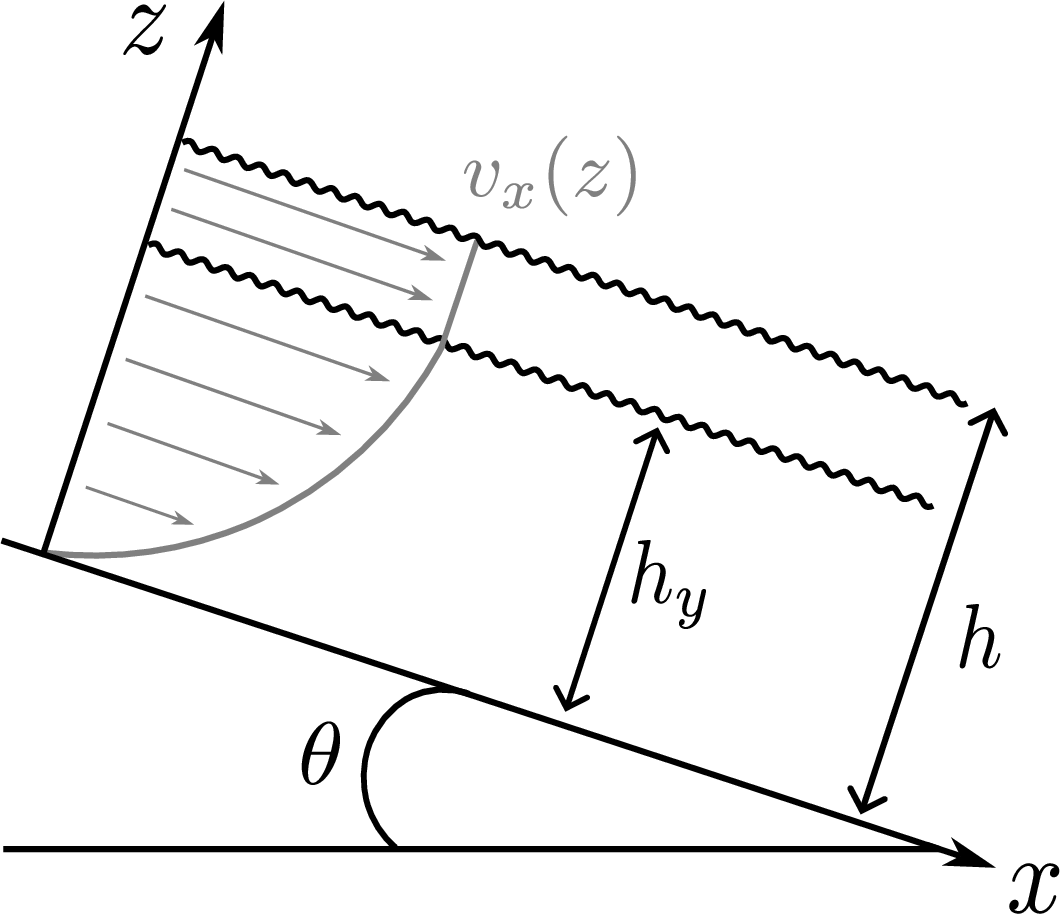}
    \caption{
    Sketch of velocity profile in a plugged surface flow.
    }
    \label{fig:pbc_setup}
\end{figure}

We consider here the flow on an inclined plane, as sketched in Fig.~\ref{fig:pbc_setup}, in order to derive steady-state velocity profiles for Peric and Duvaut--Lions fluids with both von Mises and Drucker--Prager yield with a von Mises plastic potential. 
Assuming $\sigma_{xx} = \sigma_{zz}$, we have $\tau = \rho g (h-z)\sin \theta$ and $p = \rho g (h-z)\cos \theta$.
In the figure, $h_y$ denotes the height until the start of the plug flow, and can be found as the elevation $z=h_y$  where $\tau = \tau_y$. In particular,
\begin{equation}
    h_y = h-\frac{\tau_c}{\rho g \omega}
\end{equation}
where we for convenience defined
\begin{equation}
    \omega = 
    \begin{cases}
       \sin \theta, & \text{if von Mises} \\
       \sin \theta - \mu \cos \theta , & \text{if Drucker--Prager}
    \end{cases}
\end{equation}
where we must assume $\mu > \tan(\theta)$ for flow. 
Assuming \mbox{$\dot{\gamma} = \frac{\partial v_x(z)}{\partial z}$}, the steady-state velocity profiles $v_x(z)$ resulting from the von Mises and Drucker--Prager EVP models with a von Mises plastic potential can now be derived.

Considering first the Duvaut--Lions model, Eq.~\eqref{DL_G_s}, we obtain
\begin{equation}
    \frac{\partial v_x}{\partial z}(z) = \frac{1}{t_v} \Big( \frac{\rho g \omega}{G} \Big)^{1/s}(h_y-z)^{1/s}
\end{equation}
which can be integrated to give
\begin{equation}
    v_x(z)= \frac{1}{t_v} \frac{s}{1+s}  \Big( \frac{\rho g \omega}{G} \Big)^{1/s} \Big( h_y^{\frac{1+s}{s}}-\big(h_y-z\big)^{\frac{1+s}{s}} \Big) \text{.}
    \label{ssDL}
\end{equation}
Remarkably, in the special case $s=2$, the above result has a striking resemblance to the Bagnold velocity profile which can be derived from the $\mu(I)$-rheology, 
\begin{equation}
    v_x(z) \propto \Big( h_y^{3/2}-\big(h_y-z\big)^{3/2} \Big)
    \label{bagnold}
\end{equation}
with a prefactor which depends on the material properties (density and grain size) and rheological parameters (that define the $\mu(I)$ function) in addition to the angle~$\theta$.
This is derived by, e.g., \citet[p.~232-233]{ANDREOTTIgranular2013} using the expression for $\mu(I)$ from \citet{JOPcrucial2005}.

With the Peric model, Eq.~\eqref{Peric}, we obtain the following general expression for the shear strain rate
\begin{equation}
    \frac{\partial v_x}{\partial z}(z) = \frac{1}{t_v} \bigg( \Big(\frac{(h-z)\sin \theta}{\mu(h-z)\cos \theta + (h-h_y)\omega}\Big)^{1/s} - 1 \bigg)
    \label{velprofile_peric_generalderivative}
\end{equation}
where $\mu=0$ must be replaced in the case of von Mises.
In the case of a von Mises yield, this can be integrated to give,
\begin{equation}
    v_x(z)= \frac{1}{t_v} \bigg( \frac{s}{1+s} \frac{ h^{\frac{1+s}{s}} - (h-z)^{\frac{1+s}{s}} }{(h-h_y)^{1/s}} - z \bigg)
    \label{velprofile_peric_vm}
\end{equation}
where the first term in the parentheses shares a similar $z$-dependence as Eq.~\eqref{ssDL}. 
However, with a non-cohesive Drucker--Prager yield, we obtain a linear velocity profile, regardless of the exponent $s$, 
\begin{equation}
    v_x(z)= \frac{1}{t_v} \bigg( \Big( \frac{\tan \theta}{\mu} \Big)^{1/s} - 1\bigg)z
    \label{velprofile_peric_dp}
\end{equation}
Interestingly, the velocity profiles in Eqs.~\eqref{velprofile_peric_vm} and~\eqref{velprofile_peric_dp} do not depend on any material properties other than the viscous time and viscous exponent.

\section{Numerical implementation}

\subsection{Elastic predictor -- plastic corrector}
\label{sec:numerics}
Following the approach of~\citet{SIMOalgorithms1992} in elastoplasticity, an elastic predictor -- plastic corrector scheme is employed. 
Denoting the Hencky strain $\bm{\varepsilon}^{E, t}$ predicted in an elastic step from time $T^n$, this is later corrected to
$\bm{\varepsilon}^{E, n+1}$ at time $T^{n+1}$ as
\begin{equation}
\bm{\varepsilon}^{E, n+1} = \bm{\varepsilon}^{E, t} -   \bm{l}^P \Delta t
\label{predictorcorrector}
\end{equation}
where $\Delta t =T^{n+1}-T^n$ and $\bm{l}^P$ is given in Eq.~\eqref{plasticvelgrad}.
The derivation of Eq.~\eqref{predictorcorrector} follows \citet{BONETnonlinear2008} and is given in detail in Appendix~A of \citet{BLATNYcritical2024}.

Equivalently, we can consider Eq.~\eqref{predictorcorrector} in stress space as
$\bm{\sigma}^{n+1} = \bm{\sigma}^t - \mathcal{C} : \bm{l}^P \Delta t$.
In the context of the Duvaut--Lions model, Eq.~\eqref{DL}, this can be used to immediately derive
\begin{equation}
\bm{\varepsilon}^{E, n+1} = \bm{\varepsilon}^{E, t} - b \dot{\bar{\gamma}} \frac{\partial g}{\partial \bm{\sigma}} \Delta t
\label{predictorcorrector_DL}
\end{equation}
where we defined $b = \frac{1}{1+t_v/\Delta t}$ and $\dot{\bar{\gamma}}$ is the plastic multiplier found in the equivalent rate-independent (inviscid) case.
Note that when $t_v = 0$, we get $b=1$ and Eq.~\eqref{predictorcorrector_DL} reduces to the rate-independent solution as expected. 

With a von Mises plastic potential, Eq.~\eqref{predictorcorrector} can be reduced to 
$\tau^{n+1} = \tau^t - G \dot{\gamma} \Delta t$ and $p^{n+1} = p^t$. 
Combined with the Peric model, Eq.~\eqref{Peric}, this gives 
\begin{equation}
    (\tau^t - G \dot{\gamma} \Delta t)\bigg(\frac{1}{t_v \dot{\gamma} +1}\bigg)^s - \tau_y = 0
\end{equation}
which can be solved for $\dot{\gamma}$ with iterative schemes, e.g., the Newton–Raphson method. Note that both $t_v=0$ and $s=0$ give $ \tau^{n+1} = \tau_y$ which is the rate-independent solution. 
In contrast, Eqs.~\eqref{Perzyna} and~\eqref{DL_G_s} do not provide the rate-independent solution in the limit $s\rightarrow 0$. For example, with Eq.~\eqref{DL_G_s}, the equation
\begin{equation}
    \tau^t - G \dot{\gamma} \Delta t - G(t_v \dot{\gamma})^s - \tau_y = 0
\end{equation}
must be solved for $\dot{\gamma}$.

\subsection{The volume increase in Drucker--Prager}
\label{sec:pradhana}
\begin{figure}
    \centering
    \includegraphics[width=0.7\linewidth]{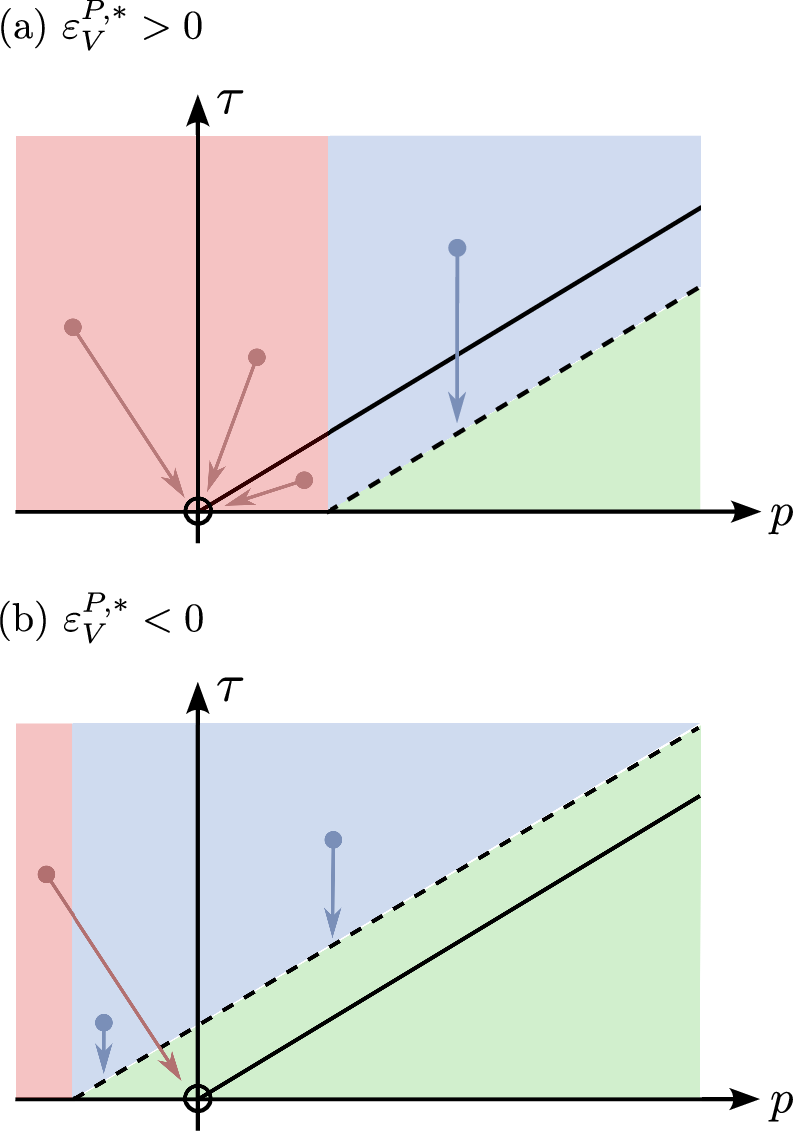}
    \caption{
    The Drucker--Prager volume correction approach of \cite{PRADHANAmultiphase2017} sketched in principal stress space. 
    The solid and dashed lines represent the original and shifted Drucker--Prager yield surface, respectively. 
    After shifting the yield, the green area represents the new elastic region while the red and blue areas are the new plastic regions, where in the latter there is no plastic volumetric deformation.  
    The arrows represent the direction of plastic flow (the return mapping).
    For simplicity, only the cohesionless case is sketched here, but the case with cohesion follows analogously.   
    }
    \label{fig:pradhana}
\end{figure}

As mentioned in Section~\ref{sec:yield}, the Drucker--Prager model is often used with a von Mises plastic potential in order to avoid excessive volume increase. 
However, the non-smoothness of this yield surface at $\tau=0$ necessitates a special plastic flow rule ensuring that, in the rate-independent case, the plastic state ends up on the yield surface, commonly, at the Drucker--Prager cone tip. 
This will induce some volumetric expansion, which may not be significant under small deformations, but may become excessive under large deformations (as will be shown in Section~\ref{sec:cliff}). 
To this end, a plastic volume correction technique similar to the one proposed by \citet{PRADHANAmultiphase2017} can be utilized. 
This technique is based on tracking a temporary plastic volumetric strain $\varepsilon_V^{P,*}$, which is initially zero, but can have both positive and negative values. 
The (rate-independent) return mapping can now be written
\begin{equation}
\bm{\varepsilon}^{E, n+1} =
\begin{cases}
    \frac{\tau_c}{dK\mu} \bm{I},  
    \text{  if  } \tr(\bm{\varepsilon}^{E,t}) > \frac{\tau_c}{K\mu} - \varepsilon_V^{P,*,n} \\       
	\bm{\varepsilon}^{E, t} - \Delta t \dot{\bar{\gamma}} \bm{n} + \frac{1}{d} \varepsilon_{V}^{P, *,n} \bm{I}, 
	\text{\ otherwise}
\end{cases}
\end{equation}
where
\begin{align}
\varepsilon_{V}^{P,*, n+1} =
\begin{cases}
	\varepsilon_{V}^{P,*, n} + \Delta \varepsilon_{V}^{P} , 
	\text{ if } \tr(\bm{\varepsilon}^{E,t}) > \frac{\tau_c}{K\mu} - \varepsilon_V^{P,*,n} \\      
	0,
    \text{\ otherwise}
\end{cases}
\raisetag{1.5\baselineskip}
\end{align}
i.e., $\varepsilon_{V}^{P,*}$ is reset to zero once we are no longer in a plastic dilating state. 
This technique can be interpreted as letting some states undergo volume reduction by projecting them to the cone tip instead of to the yield surface through the given flow rule, consequently compensating for the otherwise strict volume expansion.
This is sketched in Fig.~\ref{fig:pradhana}, highlighting the plastic flow under the case $\varepsilon_V^{P,*,n}>0$ and the case $\varepsilon_V^{P,*,n}<0$.

\subsection{The material point method}
\label{sec:mpm}

\begin{figure}[b]
    \centering
    \includegraphics[width=1\linewidth]{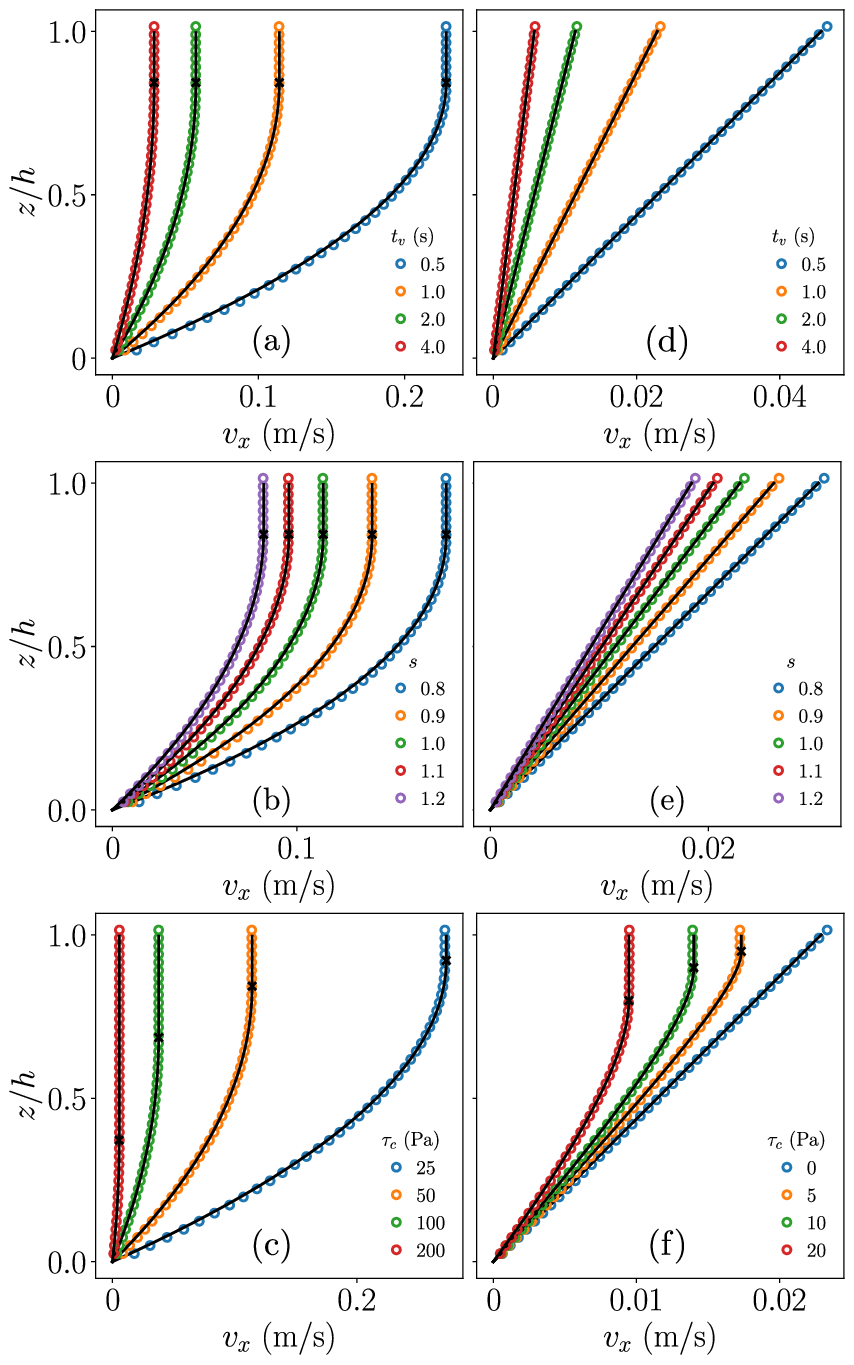}
    \caption{
      Steady-state velocity profiles from the Peric model  on a $\theta=40^\circ$ inclination, using (a)-(c) von Mises yield and (d)-(f) Drucker--Prager. 
      The circles are simulation results and the black solid lines represent the analytic solution. The black cross indicates the onset of plug flow. 
      Except when indicated in the legends, the chosen (default) parameters are $t_v = 1$~s, $s=1$, $\tau_c=50$~Pa for von Mises and $\tau_c=0$~Pa, $\mu = \tan(30^\circ)$ for Drucker--Prager.
    }
    \label{fig:peric}
\end{figure}

The constitutive laws have been implemented in the material point method~(MPM), a numerical method for solving the momentum conservation equation
\begin{equation}
	\rho(\bm{x},t) \frac{D \bm{v}(\bm{x},t)}{D t} = \nabla \cdot \bm{\sigma}(\bm{x},t) + \rho(\bm{x},t) \bm{g} \text{,}
	\label{mom_cons_eulerian}
\end{equation}
where $\rho$ is the mass density, $\bm{v}$ is the velocity,  $\bm{\sigma}$ is the symmetric Cauchy stress tensor, $D/Dt$ denotes the material time derivative and we have assumed that the only external body force present is the one resulting from the gravitational acceleration~$\bm{g}$.
Typically attributed to \cite{SULSKYparticle1994}, MPM is capable of handling large deformations and flow without mesh distortion issues and requiring no special treatment for the free surface. 
In this method, the material is discretized by Lagrangian (material) points associated with (at least) a mass~$m_p$, volume~$V_p$, velocity~$\bm{v}_p$ and strain $\bm{\varepsilon}_p$.
At every time step, the mass and velocities are interpolated to a background grid, and the momentum equation is solved on this grid.

As we are interested in flow dynamics, we focus here on an explicit MPM formulation. 
We adopt a regular grid with a grid cell size~$\Delta x$, initialized with eight material points per grid cell in three dimensions. 
Disregarding the boundary term, the discretized weak form of the momentum conservation equation can be written on the Eulerian grid nodes~$i$ for each time step~$n$ of size~$\Delta t$ as, assuming a forward Euler time discretization,
\begin{align}
\frac{   m^n_{i} \bm{v}^{n+1}_{i}  -  m^n_{i} \bm{v}^{n}_{i}    }{\Delta t}   
&=  - \sum_p V^0_{p} \bm{\sigma}(\bm{\varepsilon}_p^{E,n}) \nabla  N_{i}(\bm{x}^n_p) \nonumber \\  &\quad + m_i^n \bm{g}
\label{wf_discr}
\end{align}
where $\bm{g}$ specifies gravity, $N_{i}$ is an interpolation function and the mass on grid node $i$ is $m_i^n = \sum_p m_p N_i(\bm{x}_p^n)$. 
Since~$m_p$ remains constant in time, mass conservation is inherently satisfied.
We let $\Delta t$ be bounded by the elastic wave speed as well as the CFL condition, i.e.,
\begin{equation}
    \Delta t = \min \Bigg( C_\text{cfl} \frac{\Delta x}{\max_p(|\bm{v}_p|)}, \text{ \ }  C_\text{el} \frac{\Delta x}{\sqrt{E/\rho}} \Bigg)
\end{equation}
with the positive constants $C_\text{cfl} \leq 1$ and $C_\text{el} \leq 1$.
Boundary conditions are imposed directly on the updated grid velocities, enforcing either no-slip or frictional slip governed by a prescribed Coulomb friction. 

\begin{figure}
    \centering
    \includegraphics[width=1\linewidth]{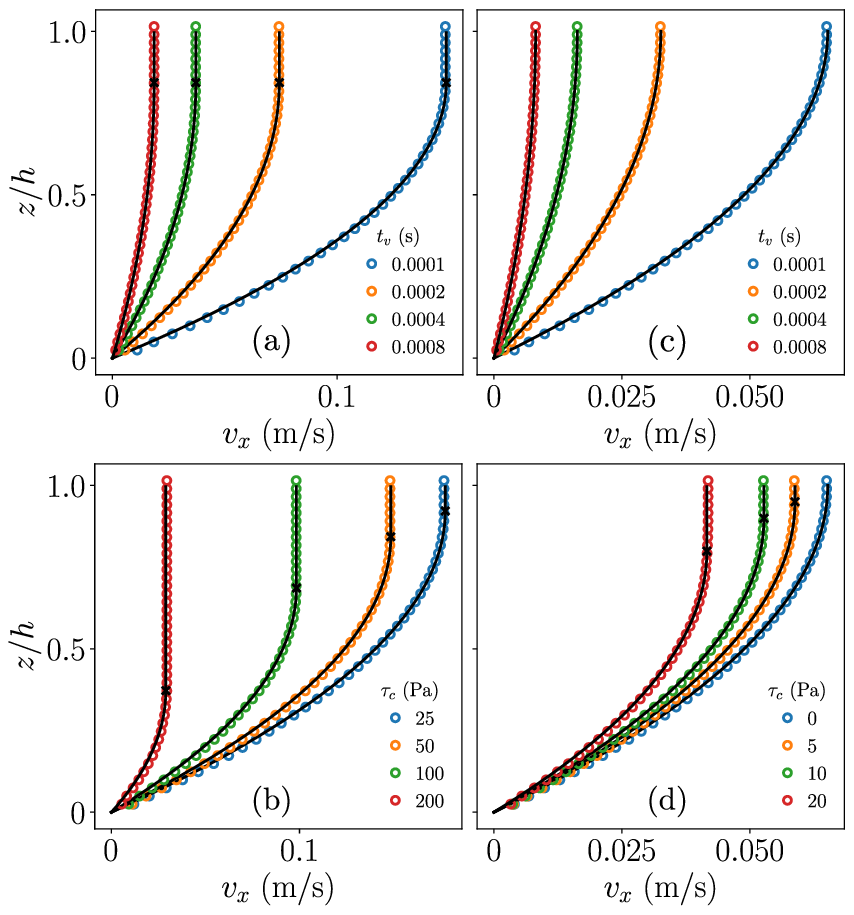}
    \caption{
      Steady-state velocity profiles from the Duvaut--Lions model on a $\theta=40^\circ$ inclination, using (a)-(b) von Mises yield and (c)-(d) Drucker--Prager. 
      The circles are simulation results and the black solid lines represent the analytic solution. 
      The black cross indicates the onset of plug flow.
      Except when indicated in the legends, the chosen (default) parameters are $t_v = 10^{-4}$~s, $s=1$, $\tau_c=50$~Pa for von Mises and $\tau_c=0$~Pa, $\mu = \tan(30^\circ)$ for Drucker--Prager.
    }
    \label{fig:dl}
\end{figure}

\begin{figure}
    \centering
    \includegraphics[width=1\linewidth]{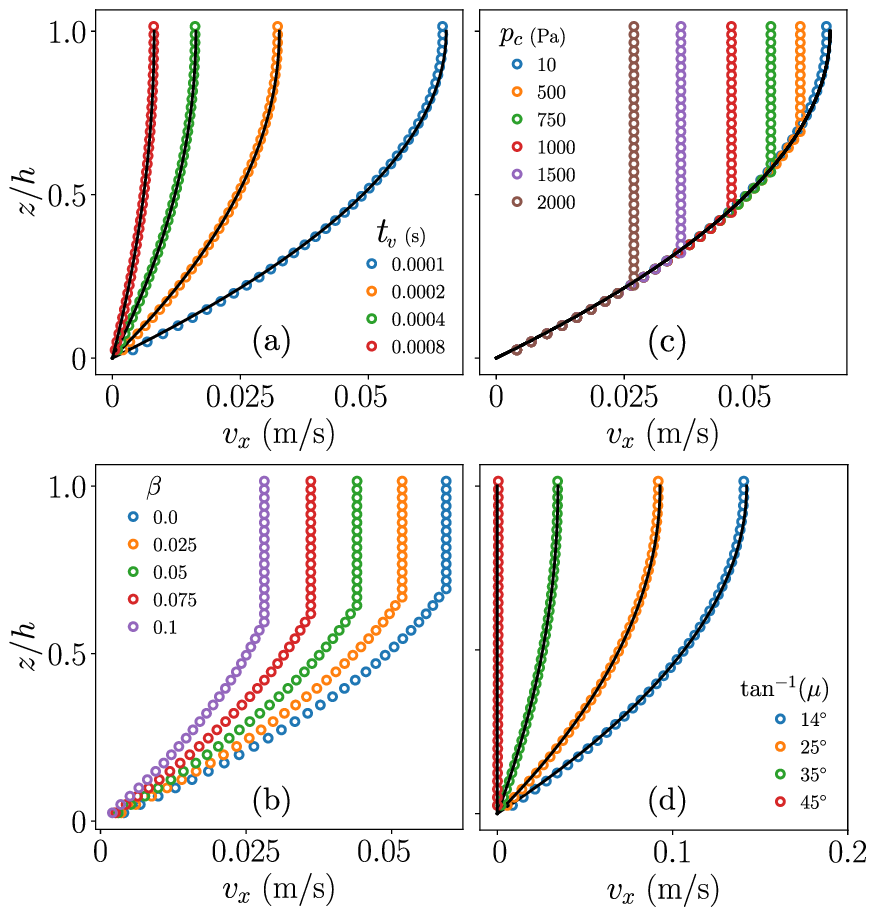}
    \caption{
      Steady-state velocity profiles from the Duvaut--Lions model using the modified Cam--clay on a $\theta=40^\circ$ inclination. 
      The circles are simulation results and the black solid lines represent the cohesionless analytic solution assuming all states lie on the critical state line. 
      Except when indicated in the legends, the chosen (default) parameters are $t_v = 10^{-4}$~s, $s=1$, $\mu = \tan(30^\circ)$, $\beta=0$, $p_c = 10$ Pa and $\xi = 10^3$.
    }
    \label{fig:mcc}
\end{figure}

Initial critiques of~MPM raised concerns about cell-crossing instabilities and numerical dissipation. 
In recent years, the~MPM community has tackled these problems through the use of B-spline interpolation functions~\citep{STEFFENanalysis2008} as well as the adoption of affine transfer schemes such~APIC~\citep{JIANGangular2017} and subsequently~AFLIP~\citep{FEIrevisiting2021} to reduce dissipation.
Further details about~MPM and the implementation used in this work can be found in \cite{BLATNYmatter2025a}.

Although we have opted for~MPM in this study, other implementations of the same constitutive framework can be accomplished in other methods with similar features, such as the Smoothed Particle Hydrodynamics~(SPH), the Finite Element Method with Lagrangian Integration Points~(FEMLIP) and the Particle Finite Element Method~(PFEM), which all have been used for viscoplastic simulations of yield-stress fluids (e.g., \citet{CHAMBONnumerical2011, AHONGUIOflow2016, FRANCI3d2019, GUEYEnumerical2021}).
Each method comes with its own advantages and disadvantages. 
For example, MPM demands extra memory due to its background grid, FEMLIP requires a treatment of the free surface, SPH is associated with expensive particle neighbor search and challenging handling of boundary condition and in PFEM frequent remeshing is necessary.

\section{Examples}

\subsection{Steady-state flow}

With the setup sketched in Fig.~\ref{fig:pbc_setup}, using periodic boundary conditions in the $x$-direction, we investigate whether the numerical implementation can retrieve the steady-state velocity profiles found in Section~\ref{sec:profiles}. 
Using the Peric model, Eq.~\eqref{Peric}, and  the Duvaut--Lions model, Eq.~\eqref{DL_G_s}, this is shown in Figs.~\ref{fig:peric} and~\ref{fig:dl}, respectively, using both a von Mises yield (a-c) and Drucker--Prager yield (d-e).
In all cases the elastic parameters $E=1$~MPa and $\nu=0.3$ are used, the inclination angle is $\theta=40^\circ$ and the velocity profile is measured at $t=0.5$~s after starting from rest.

There is no general analytical solution for the steady-state velocity profile with a modified Cam--clay yield. 
However, with a sufficiently small initial value of the initial compressive strength~$p_c$ and a sufficiently large value of the hardening parameter~$\xi$, we propose that the plastic states will reside around the critical state line determined by~$\mu$, and, as such, we expect the solution to be equivalent to that of Drucker--Prager. 
Figs.~\ref{fig:mcc}a and~d show that, indeed, the Drucker--Prager solution can be retrieved for various values of~$t_v$ and~$\mu$.
Moreover, Figs.~\ref{fig:mcc}b and~c show the departure from this solution when~$p_c$ and~$\beta$ are increased, providing a control on the emergence of plugged surface flow.

While the overstress approach ensures that steady-state flow can be induced, the purely inviscid model ($t_v = 0$) does not produce steady-state flow, rather it will accelerate indefinitely if $\mu < \tan \theta$ or come to rest if $\mu > \tan \theta$. 
As such, the overstress approach, similar to the $\mu(I)$-rheology, provides a means of regularizing this instability of the inviscid Drucker--Prager model.

\subsection{Dam break}
\label{sec:dambreak}

\begin{figure}[t]
    \centering
    \includegraphics[width=0.95\linewidth]{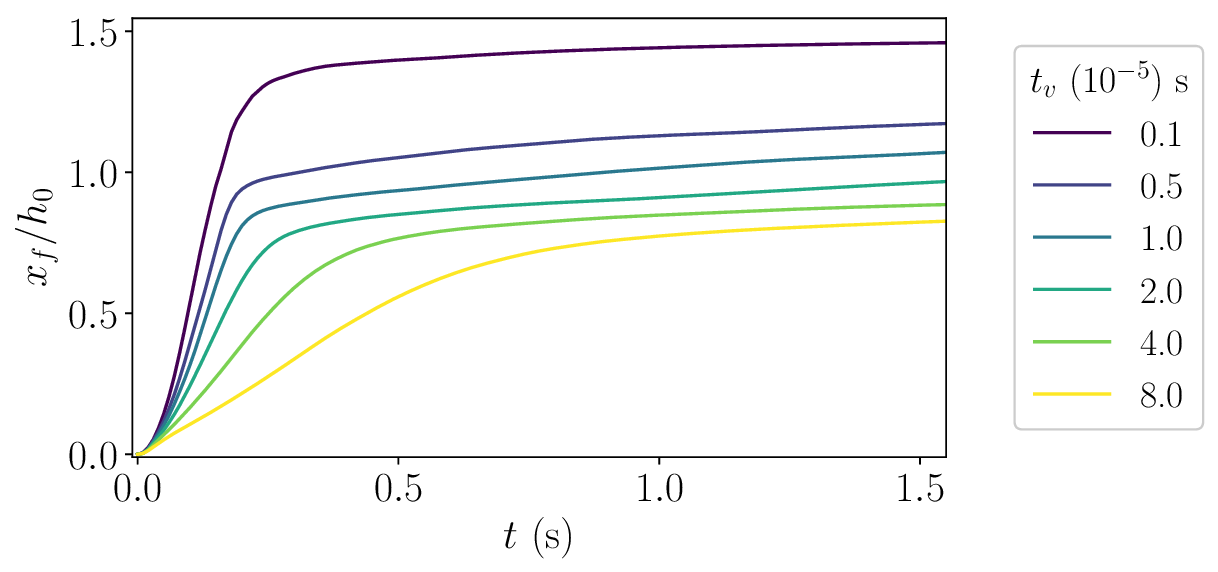}
    \caption{
    Dam break, initially $7 \times 5$ cm, released on an inclination~$\theta=10^\circ$ using modified Cam--clay with parameters $\mu = \tan(30^\circ)$, $\beta=0$, $p_c = 10$~Pa, $\xi = 10^2$, $E=1$~MPa and $\nu=0.3$. 
    The plotted runout distance is here normalized by the initial height~$h_0$. 
    }
    \label{fig:collapse_xf}
\end{figure}

\begin{figure}
    \centering
    \includegraphics[width=0.95\linewidth]{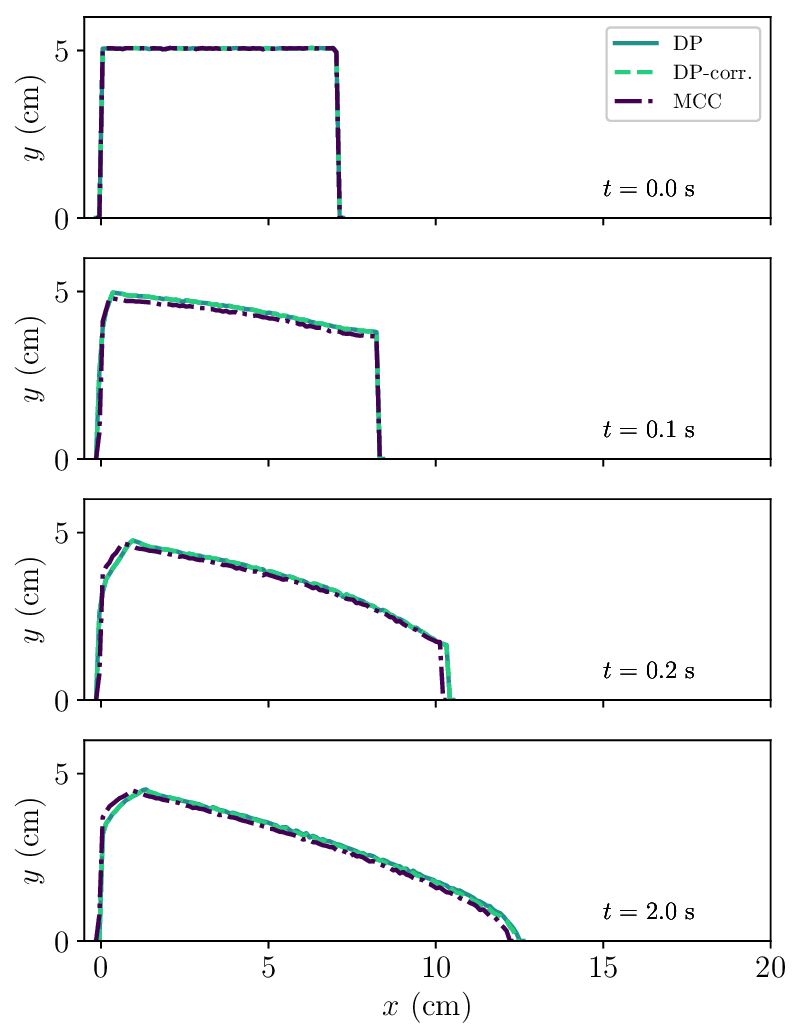}
    \caption{
    Dam break, initially $7 \times 5$ cm, released on an inclination~$\theta=10^\circ$. Three cases corresponding to Drucker--Prager~(DP), Drucker--Prager with volume correction~(DP-corr.) and modified Cam--clay~(MCC) are compared, with parameters $\mu = \tan(30^\circ)$, $\beta=0$, $p_c = 10$~Pa and $\xi = 10^2$, all using a viscous time~$t_v=2\cdot10^{-5}$. 
    The elastic parameters $E=1$~MPa and $\nu=0.3$ are used.
    }
    \label{fig:collapse_time_M30}
\end{figure}

The preceding example demonstrated that modified Cam--clay can reproduce the solution with Drucker--Prager for infinite shear flow. 
To investigate this beyond this idealized setting, we consider the case of a dam break released on a no-slip surface, inclined an angle $\theta = 10^\circ$ from the horizontal. 
Similar setups, with various inclinations, has been used by various researchers to study the behavior of yield-stress fluids, e.g.,~\cite{COUSSOTdetermination1995, ANCEYdambreak2009, MANGENEYerosion2010}.

With the choice of an internal friction $\mu > \tan(\theta)$, the dam break will experience a finite runout distance $x_f$ along the inclined plane.
This runout decreases as a function of viscosity, e.g., as illustrated in the example provided in Fig.~\ref{fig:collapse_xf} using the viscous modified Cam--clay model.
In fact, with comparable parameters, there is no significant difference between the Drucker--Prager and modified Cam--clay models in the runout or collapse of the dam break, as shown in Fig.~\ref{fig:collapse_time_M30}.
This figure also includes the corrected Drucker--Prager model from Section~\ref{sec:pradhana}.

\begin{figure}
    \centering
    \includegraphics[width=0.9\linewidth]{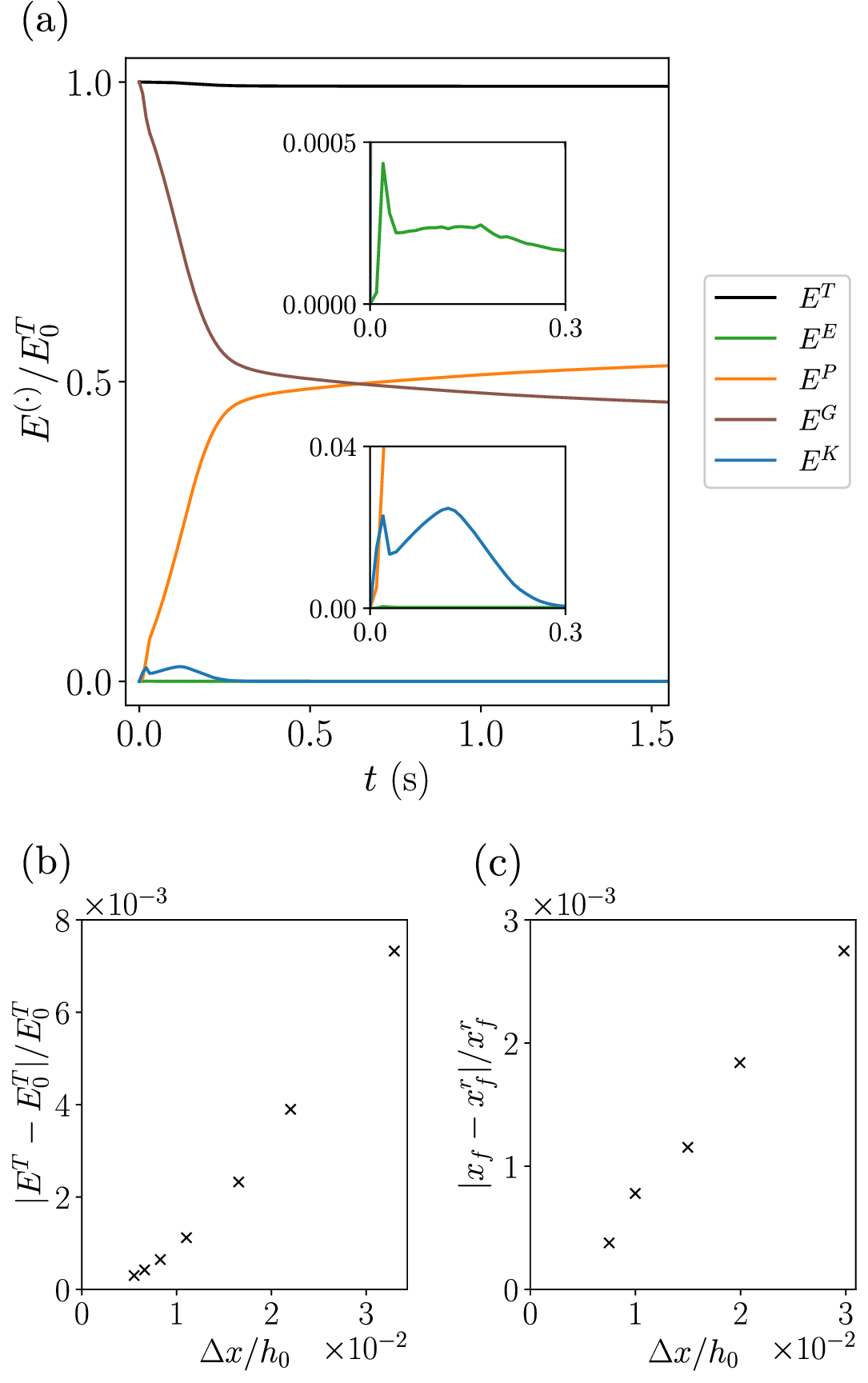}
    \caption{
    Energy and convergence of the MCC dam break from Fig.~\ref{fig:collapse_time_M30}. 
    (a) Normalized total~($E^T$), elastic~($E^E$), plastic~($E^P$), potential~($E^G$) and kinetic  energy~($E^K$) during the dam break, where the two inset plots show zoomed views on the elastic and kinetic energies, respectively. 
    (b) The relative total energy loss is reduced with finer resolution~$\Delta x$.
    (c) The runout distance~$x_f$ converges with finer resolution relative to a reference solution~$x_f^r$.
    }
    \label{fig:collapse_convergence}
\end{figure}

Fig.~\ref{fig:collapse_convergence} shows that the total energy in the system is approximately conserved by the numerical scheme (Fig.~\ref{fig:collapse_convergence}a) and that the small energy loss decreases with a finer spatial resolution (Fig.~\ref{fig:collapse_convergence}b). 
The runout distance also converges with a finer resolution, as shown in Fig.~\ref{fig:collapse_convergence}c. 
In Fig.~\ref{fig:collapse_convergence}a, the various contributions to the total energy are included, showing that the loss of potential energy is mainly converted into plastic dissipation, highlighting the minor contribution of the kinetic energy in the initial collapse. 

The corrected Drucker--Prager model provides a similar result as the standard Drucker--Prager model, with only a very slightly reduced runout, which can be explained by the minimal volumetric expansion occurring in this example. 
Therefore, in the next example, we consider a case that displays a stronger volumetric deformation.

\subsection{Stretching flows}
\label{sec:cliff}

\begin{figure*}
    \centering
    \includegraphics[width=1\linewidth]{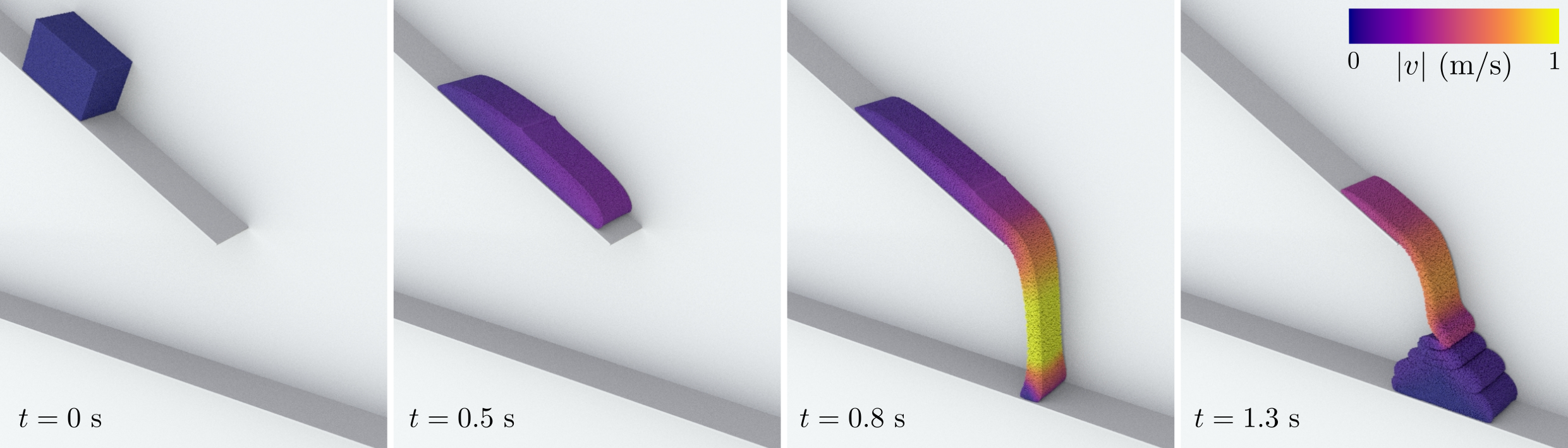}
    \caption{
    A material flowing over a "cliff" and impacting the ground, as described in from Section~\ref{sec:cliff}, here simulated with the Drucker--Prager model. 
    Initial dimensions of the material are $7 \times 5 \times 4$ cm. 
    The parameters used are $E=1$~MPa, $\nu=0.3$, $\mu = \tan(30^\circ)$, $\beta=0$, $p_c = 200$~Pa, $\xi = 10^2$ and $t_v=2 \cdot 10^{-5}$. 
    The material is colored according to its speed. 
    }
    \label{fig:cliff_render}
\end{figure*}

\begin{figure}[b]
    \centering
    \includegraphics[width=1\linewidth]{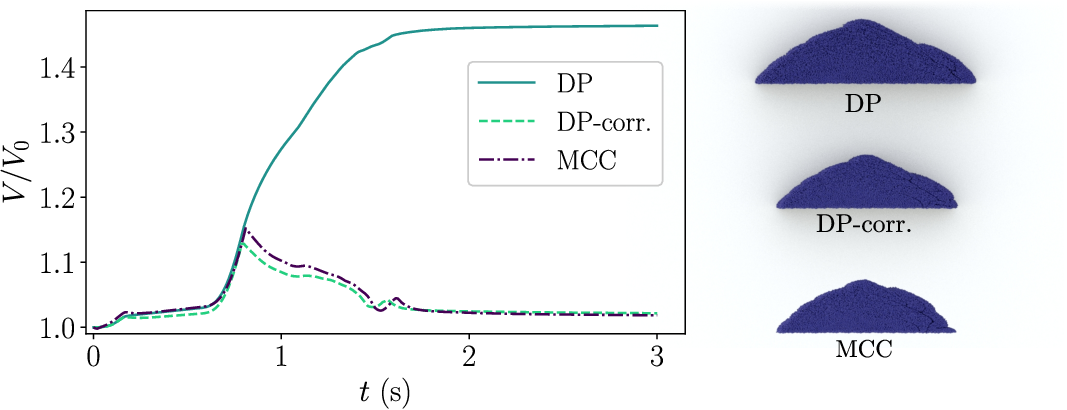}
    \caption{
    The change in total volume~$V$ (normalized by the initial volume~$V_0$) during the simulations from Section~\ref{sec:cliff}, from the initial collapse to steady inclined flow before the free fall and the following pile-up on the ground.
    On the right are the final states at $t=3$~s, using Drucker--Prager~(DP), Drucker--Prager with volume correction~(DP-corr.) and modified Cam--clay~(MCC).
    }
    \label{fig:cliff_volume}
\end{figure}

In order to further demonstrate the differences between the corrected and uncorrected Drucker--Prager model, as well as the modified Cam--clay model, which all carry the notion of an internal friction parameter~$\mu$, we consider here a material released on an inclined plane as in the previous dam break simulations, here with an inclination of $\theta = 30^\circ$. 
This flow is subsequently stretched from falling off the end of this plane, dropping vertically before impacting the ground, consequently inducing a folding of the material.
The simulation with Drucker--Prager is illustrated in Fig.~\ref{fig:cliff_render}.
We consider a frictional boundary condition, for simplicity with a Coulomb friction $\mu_c=\mu=\tan(\theta)$, and we allow the material to separate from the boundary.
Simply looking at the final rest state of the piled up material from the simulations using Drucker--Prager, the corrected Drucker--Prager and modified Cam--clay, we see in Fig.~\ref{fig:cliff_volume} that the Drucker--Prager has gained a significant amount of volume. 
In particular, Drucker--Prager induces an increase of the volume of the flowing material by around 50\%. 
Surprisingly, the corrected Drucker--Prager and modified Cam--clay give almost identical volumetric behavior, increasing slightly during the stretching phase (until about $t=0.9$~s after release), before compacting as the material impacts the ground and starts folding.

\section{Discussion and conclusions}
In this work, we have demonstrated, through representative examples, how the choice of yield surface (von Mises, Drucker--Prager and modified Cam--clay) influences the behavior of EVP fluids. 
Rate-dependent plasticity was considered through the overstress approach of Perzyna and Duvaut--Lions, between which we have highlighted their differences as well as their general equivalence under certain model choices. 
Moreover, we analytically derived velocity profiles for idealized shear flows and showed that the Bagnold velocity profile for granular flow can be considered a special case of the Duvaut--Lions model. 
We chose~MPM as the numerical scheme in which we implemented the proposed finite strain framework, however, it must be emphasized that this framework could also be implemented in other schemes.  
For example,~\cite{AHONGUIOflow2016} provides a different EVP framework in FEMLIP, which, like the framework proposed here, does not require any regularization of the discontinuity due to the solid-fluid transition. 
    
Drucker--Prager should generally not be accompanied by an associative flow rule due to excessive volume increase. 
As demonstrated, even with a von Mises plastic potential (i.e., a non-associative flow rule), Drucker--Prager can cause excessive volume increase during large deformations, due to the special treatment of the yield surface singularity (typically treated with a cone tip projection).
While this volumetric expansion is typically negligible in small-strain solid mechanics, it must be carefully considered when modeling fluid flows. 
To this end, we have shown that the correction of~\citet{PRADHANAmultiphase2017}, initially proposed in the computer graphics community, can help reduce excessive volume increase. 
    
Alternatively, the (viscous) modified Cam--clay model can be used to retrieve the solution of the (viscous) Drucker--Prager, without any problems of excessive volume increase. 
This was shown for shear flow and dam break problems. 
However, it is important to highlight that Cam--clay models can also be used (with other parameter choices) to produce very different results than Drucker--Prager and, unlike Drucker--Prager, has great potential in modeling compressible flows. 
The compressible behavior can be specified through the hardening law, for which we here chose a relatively simple relationship given by Eq.~\eqref{hardening_exp}.
We also restricted ourselves to relatively large values of the hardening parameter~$\xi$ in order to induce a hardening behavior that promoted stress states to rather quickly align with the critical state. 
This was a deliberate choice in order to critically compare modified Cam--clay with Drucker--Prager. 
The compressible properties of Cam--clay, especially for solid behavior, are beyond the scope of this contribution, and the reader is referred to classical geomechanics literature (e.g., \cite{WOODsoil1991, PUZRINconstitutive2012} and others) for the general characteristics of this model. 

We have not studied the influence of elastic properties and we have restricted ourselves to one (hyper)elastic model. 
The chosen elastic model, Hencky's elasticity model, has desirable features which are further discussed in \citet{BRUHNSselfconsistent1999, XIAOhenckys2002} and makes the numerical implementation particularly easy.
Although the plastic properties are usually dominant during flow, the elastic properties may nevertheless be highly relevant and may also be rate-dependent. 
While we have considered yield surfaces that broadly represent the pressure-independent, frictional and fully capped families, our selection is not exhaustive. 
For example, the Tresca model, which is pressure-independent like von Mises, the Mohr--Coulomb model, which bears similarities to Drucker--Prager, the capped Drucker--Prager model~\citep{DRUCKERsoil1957}, which extends Drucker--Prager to yield also under compression similar to modified Cam--clay, as well as more modern models such as Matsuoka–-Nakai~\citep{MATSUOKAstressdeformation1974, HOULSBYgeneral1986} and Nor-sand~\citep{JEFFERIESnorsand1993} may also be considered. 
The (solid) properties of these models may be found in classical geomechanics literature.

\section*{Acknowledgments}
The authors thank Johan Gaume and Guillaume Chambon for fruitful discussions. L.B. gratefully acknowledges financial support from the Swiss National Science Foundation~(SNSF) through grant number~P500PT\_230265.

\section*{Data availability statement}
All models presented here are implemented and freely available in the open-source MPM software \textit{Matter}~\citep{BLATNYmatter2025, BLATNYmatter2025a}.

\section*{CRediT authorship contribution statement}
\textbf{Lars Blatny}: Conceptualization of this study, Methodology, Software, Validation, Formal analysis, Investigation, Writing - Original Draft, Visualization, Funding acquisition. \textbf{Alexandre Pellet}: Methodology, Software, Writing - Review \& Editing.

\end{document}